\documentclass[a4paper,12pt,prb,amsmath,amssymb,floatfix,preprint]{revtex4}

\usepackage{widetable}
\usepackage[usenames,dvipsnames]{color}
\usepackage[pdftex]{graphicx} 
\DeclareGraphicsExtensions{.pdf}
\usepackage{color}
\usepackage{array}
\newcolumntype{C}[1]{>{\centering\arraybackslash}m{#1}}
\begin{document}
\title{Heusler compounds -- how to tune the magnetocrystalline anisotropy}
\author{H. C. Herper}
\email{heike.herper@physics.uu.se}
	\affiliation{Department of Physics and Astronomy, Uppsala University, Box 516, 751 20 Uppsala, Sweden}	
\begin{abstract}
Tailoring and controlling  magnetic properties is an important factor for materials design. Here, we present a case study for Ni-based Heusler compounds of the type Ni$_2$YZ  with Y = Mn, Fe, Co and  Z = B, Al, Ga, In, Si, Ge, Sn based on first principles electronic structure calculations. These compounds are interesting since the materials properties can be quite easily tuned by composition and many of them possess a non-cubic ground state being a prerequisite for a finite magnetocrystalline anisotropy (MAE).  We discuss  systematically the influence of doping at the Y and Z sublattice as well of lattice deformation on the MAE. We show that in case of Ni$_2$CoZ  the phase stability and the MAE can be improved using quaternary systems  with elements from group 13 and 14 on the Z sublattice whereas changing the Y sublattice occupation by adding Fe does not lead to an increase of the MAE. Furthermore, we studied the influence of  the lattice ratio on the MAE. Showing that small deviations can lead to a doubling of the MAE as in case of Ni$_2$FeGe. Even though we demonstrate this for a limited set of systems the findings may carry over to other related systems.  \end{abstract}
\maketitle
\section{Introduction}
\begin{figure}[b]
\centering
\includegraphics[width=.9\columnwidth]{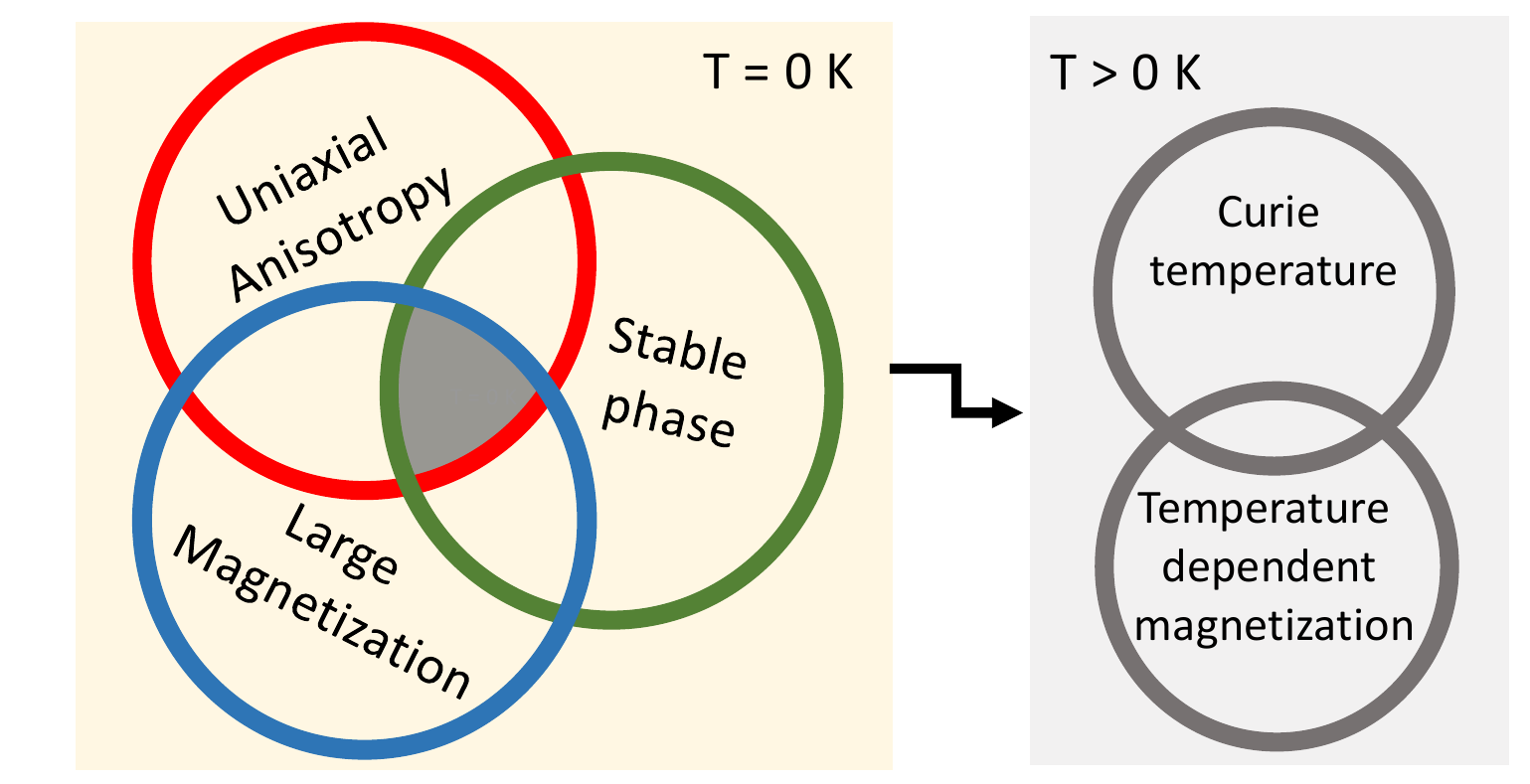}
\caption{Computational design of magnetic materials: Basic properties which can be obtained from first principles calculations ($T = 0$\,K) and temperature dependent quantities which can be determined from additional atomistic modeling.}
\label{fig:magn-prop}
\end{figure}
The demand for new magnetic materials is bigger than never before. Applications are very diverse and  comprise permanent magnets for cars and wind turbines\cite{}, actuators\cite{}, memory devices\cite{}, as well as for magnetic cooling\cite{} where different applications also demand for different technical specifications of the magnetic material.  
An eminent goal is to identify new materials for magnetic applications. A resource saving and often faster way compared to experiments is computational materials design using {\em ab initio} methods or atomistic modeling \cite{}. In principle two different routes exist: high throughput data mining to find unknown phases or optimization and modification of known structures. In practice often high throughput studies are carried out with certain constraints on the structure or other  properties. A recent example can be found in Ref.\,\onlinecite{Sanvito:17} where this technique has been used to find new magnetic Heusler compounds.
 In the latter case the challenge is to find out on which screw to turn to optimize all relevant properties.  From first principles point of view the focus is on three properties as depicted in Fig.~\ref{fig:magn-prop}: the stability of the phase, a ferromagnetic phase with suitable high  magnetization, and last but not least the magnetocrystalline anisotropy (MAE) which is crucial for the magnet. As depicted in Fig.~\ref{fig:magn-prop} the properties are related to each other, e.g. a large magnetization without a suitable large MAE will not result in a hard magnet because the coercivity would be too small. Adding atomistic model calculations further properties such as the Curie temperature can be predicted. However, here the focus is on the basic properties shown on the left side of Fig. \ref{fig:magn-prop}. Using Ni-based Heusler compounds as model system we discuss the influence of mechanical deformation, alloying, and electronic structure on the  basic magnetic properties. 
Ni-based Heusler compounds  are known to show in certain compositions a tetragonal instability\cite{} which makes them an ideal test system even if the expected Curie temperatures are too low for high performance magnets. During the last decades Heusler alloys have been discussed as possible candidates for different magnetic applications, e.g. halfmetallic Co$_2$FeSi for spintronics applications~\cite{Meinert:12,Sagar:14,Wurmehl:07}, as rare earth (RE) free permanent magnets~\cite{Kiss:13,Nordblad:15}, or Ni-based actuators~\cite{Ayuela:99,Kalimullina:14} and magneto-caloric materials~\cite{Ren:15,Buchelnikov:11}, because of their magnetic and electronic properties can be quite easily tuned by composition. \cite{Graf:13,Winterlik:12,Buchelnikov:10,Ke:10}  Depending on the application the key properties are a high spin polarization,  a large magneto crystalline anisotropy (MAE), a high Curie temperature or a large magnetic shape memory effect.\cite{Entel:11a} It has been shown that Co based Heusler alloys such as Co$_2$FeSi and Co$_2$MnSi are half metallic ferromagnets  with magnetic moments following the Slater-Pauling curve and very high Curie temperatures\cite{Wurmehl:05, Galanakis:06}, Ni based systems, e.g. Ni$_2$MnGa or off-stoichiometric Ni-Mn-Z (Z = Sn, Sb, In) are well known for their shape memory behavior.\cite{Buchelnikov:10,Siewert:11,Krumme:15,Klaer:14} Recently the MAE on Mn based Heusler alloys has been discussed in view of their suitability  for spin transfer torque applications\cite{Winterlik:12}. These systems are ferrimagnetic with a  small net moment of 1-2$\,\mu_{\rm B}$ but MAE values up to 1meV/f.u. Furthermore, the search for new rare earth free or lean ferromagnets  has become  highly topical since permanent magnets with high MAE are needed en masse, such that cheap and abundant alternatives to the critical RE magnets are needed.  Here we focus on ferromagnetic Heusler systems and  in particular on their magnetic properties. The goal is to reveal routes to improve the magnetic properties -- in particular the MAE. Even though the absolute values might not  reach the MAE of RE \cite{} or Pt \cite{} containing materials and the Curie temperatures are expected to be lower than for high performance magnets the properties might still be comparable to bonded magnets and ferrites but Heusler alloys have a huge potential due to the easy tuning by composition and they are comparably cheap.
Heusler alloys  X$_2$YZ usually crystallize in L2$_1$ structure with point group 225 (Fm$\bar{3}$m symmetry). It consists of four interpenetrating fcc lattices, see Fig.~\ref{fig:struc}. In ordinary magnetic Heusler alloys the sub lattices A and C are occupied by  the metal X whereas another transition metal Y sits on the B site. The last sub lattice D is occupied by a main group element Z.  In some cases the occupation of the B and C sub lattice is interchanged which leads to a reduction in symmetry, i.e. point group 216 (F$\bar{4}$3m symmetry) and the so called inverse Heusler structure. Which structure is preferred depends in general on the choice of the X and Y element. For systems with the X element having a higher atomic number than the Y element the normal L2$_1$ structure is assumed to be the most stable structure whereas inverse ordered compounds appear for the opposite case~\cite{}. However, we will show that the rule is not strictly followed by Ni$_2$YZ compounds.
\\
Ni based Heusler alloys have attracted quite some interest in view of ferromagnetic shape memory alloys (FSMA)~\cite{}. They tend to  possess a tetragonal instability, i.e. they undergo a martensitic transformation from the high temperature cubic phase to a low temperature tetragonal distorted or in some cases to a modulated phase (e.g. 5M,14M)~\cite{Uijttewaal:09,Endo:11,Tanaka:04}. Since a tetragonal instability is fundamental for shape memory alloys and magneto caloric systems quite some effort has been done to find tetragonally distorted Heusler systems.\cite{}  Furthermore, the tetragonal distortion gives rise to a MAE and its dependence on the occupancy of the Y lattice and the choice of the main group element is discussed here.  The appearance and stability of the tetragonal state depends on the choice of the Y metal and the main group element. Here we study N$_2$YZ  Heusler alloys with Y = Mn, Fe, Co and  Z being a III or IV main group element and the trends in the magnetic properties especially the MAE depending on the Y and Z element  within different  DFT methods.
Even though the calculated MAE  does not reach the high  values of Pt containing Heusler alloys we show that reasonable values due to lattice deformation and out-of plane orientation of the easy axis can be achieved without 5d or RE elements and they  bear certain potential for magnetic applications.
\\
After a brief description of the computational methods in Section \ref{sec:methods} the stability of the Heusler compounds is discussed in Sec.~\ref{sec:phase}. Section~\ref{sec:magn} focusses on the MAE and the magnetic moments pointing out trends and possibles routes to increase the MAE followed by concluding remarks in Sec.~\ref{sec:conclusion}.
\section{Methods}\label{sec:methods}
\begin{figure}[b]
\centering
\includegraphics[width=.9\columnwidth]{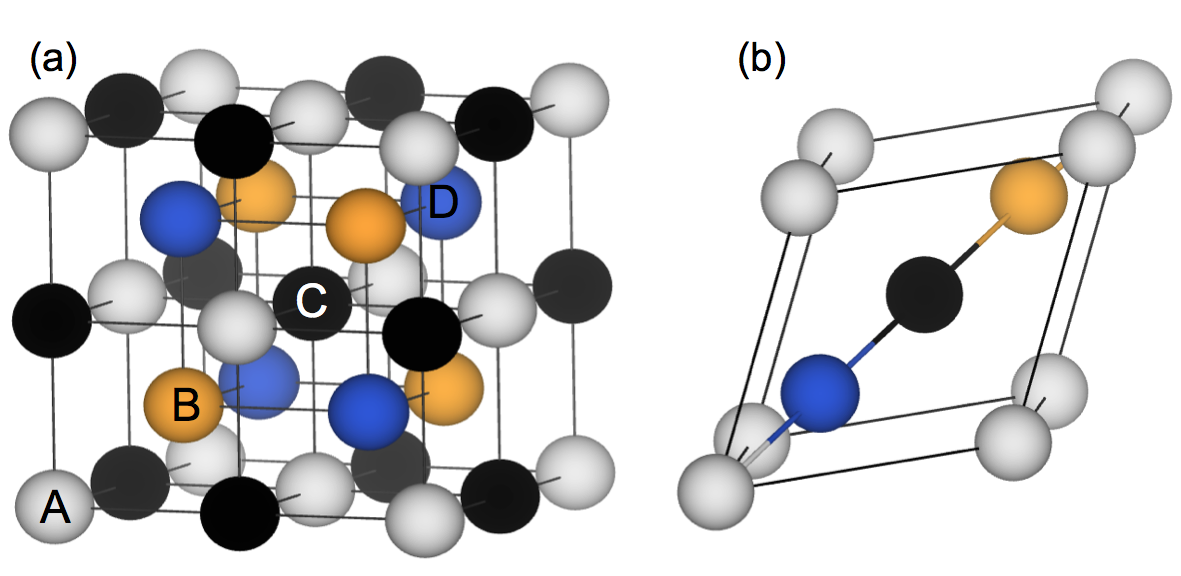}
\caption{(a) Primitive cell of the cubic L2$_1$ Heusler structure and the corresponding unit cell (b). In ordinary Heusler alloys with space group 225 the sublattices A an C are occupied by material X, B hosts the Y metal, and the main group element Z is located on the D sub lattice. For the inverse Heusler structure (Hg$_2$CuTi) the occupation of the sub lattices B and C is inverted.}
\label{fig:struc}
\end{figure}
The electronic and magnetic structure of Ni$_2$YZ Heusler compounds has been investigated within the VASP code~\cite{Kresse:96, Kresse:94} employing the projected augmented wave (PAW) potentials~\cite{Bloechl:94} and the approximations of Perdew, Burke, and Ernzerhof~\cite{Perdew:96} for the exchange correlation functional. The calculations of the stoichiometric systems have been  preformed within the 4 atomic primitive cell, see Fig.~\ref{fig:struc} (b). 
Systems have been relaxed to forces below 0.01 eV/\AA.  For calculations within the primitive cell a 17$^3$ k-point mesh has been used for structural optimization. Quaternary systems and systems with site disorder have been treated in a 16 atomic cubic cell using a k-point mesh of 13$^3$.  The c/a variation has been performed with the same accuracy but the volume and the shape of the unit cell have been fixed if not stated otherwise.
\\
The MAE calculations have been performed within the full potential linearized augmented plane wave (LMTO) approach employing the RSPt code~\cite{RSPt} whereby the  optimized structures from the previous VASP investigations serve as  input for the LMTO code and no further structural optimization has been done. For consistency the same functional for exchange and correlation has been chosen as before. A mesh of  54$\times$54$\times$54 points has been used to calculate the MAE from the 4 atomic unit cells after a  k-point convergence check. In case of off-stoichiometric systems and 16 atomic super cells slightly smaller meshes were used. Within the RSPt code the wave functions are expanded to $l_{\rm max} =8$. 
\section{Structural stability}\label{sec:phase}
\subsection{Cubic vs. tetragonal}
\begin{figure}[tbh]
\centering
\includegraphics[width=.9\columnwidth]{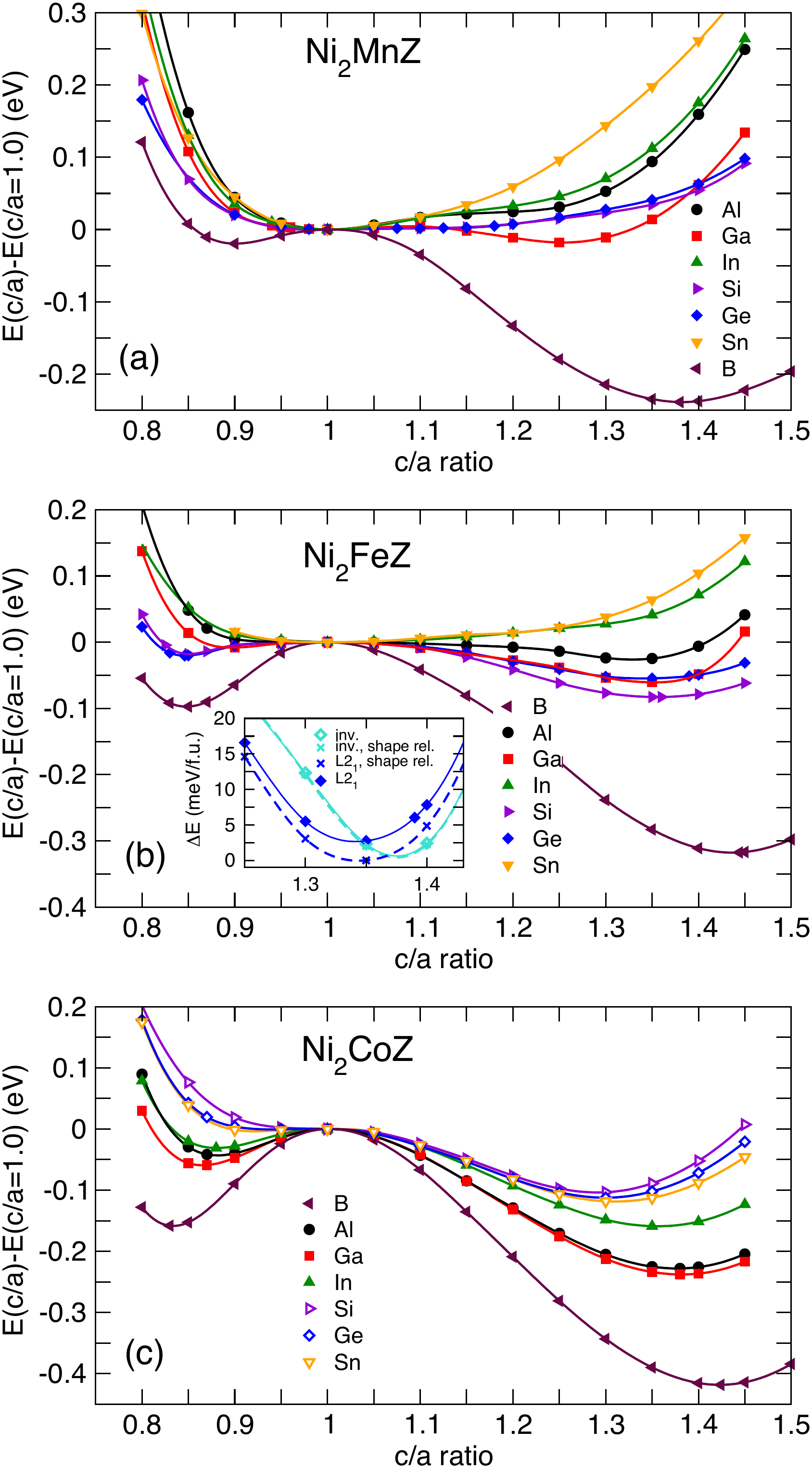}
\caption{Energy as function of the c/a ratio for Ni$_2$YZ Heusler alloys for Z elements with three (B, Al, Ga, In) and four (Si, Ge, Sn) valence electrons. Filled symbols denote normal Heusler structure X$_2$YZ. Systems which crystallize in the inverse Hg$_2$CuTi structure are marked by open symbols. Please denote the different scales on the y axis for Y = Mn (a).  The inset in (b) shows the E(c/a) for L2$_1$ and inverse ordered Ni$_2$FeGe in the vicinity of the global minimum.
\label{fig:coa}}
\end{figure}

Heusler alloys with Mn on the Y site and a main group element from group III or IV on the Z site turned out to be regular ordered alloys with Fm${\bar 3}$m symmetry in the cubic phase. However besides the well known Ni$_2$MnGa only Ni$_2$MnB possesses a tetragonal instability with c/a = 1.38 and a local minimum at c/a = 0.9 which was also found in literature\,\cite{Qawasmeh:12}, see Fig.\ref{fig:coa}(a). All other investigated Ni$_2$MnZ alloys have a cubic ground state. The only controversially discussed system is Ni$_2$MnGe which in agreement with experimental findings by Oksenenko {\em et al.}\cite{Oksenenko:06} turns out to be cubic from the present calculations using PAW potentials but Luo {\em et al.} observed tetragonal ground state from DFT calculations with US pseudo potentials\,\cite{Luo:13}. In addition calculations of the phonon spectra indicate a softening of the TA$_2$ mode which also hints to an instability of the cubic phase. A close look to E(c/a) curve reveals that there is an indication of a shallow local minimum around c/a = 1.05 being only 1.22\,meV higher in energy than the ground state. Hence small distortions or different choice of potentials may be sufficient to reverse the order if the local minima and stabilize the tetragonal phase. 
\\
In agreement with previous investigations\,\cite{Gillessen:10,Qawasmeh:12,Kreiner:14}  a tetragonal ground state is observed for nearly all investigated systems if the Y site is occupied with Fe. Exceptions are Ni$_2$FeIn and Ni$_2$FeSn which remain cubic, see Fig.\ref{fig:coa}(b). In case of Sn an indication for a saddle point can be spotted at c/a = 1.2 but no real minimum appears for this system. The global minimum for the tetragonal systems occurs at c/a = 1.35, only for the B containing system it turns out to be larger than $\sqrt 2$. Nearly all investigated Ni$_2$FeZ prefer the ordinary Heusler structure, however in some cases the energy differences between the ordinary and inverse phase are extremely small, see Fig.\ref{fig:emix}(a) and for a summary of all c/a ratios including values from literature we refer to Tab.\,\ref{tab:coa}.   A prominent example is the Ni$_2$FeGe system in which the inverse ordered structure is only 0.2 meV in lower energy than the L2$_1$ ordered phase. A inverse ordered ground state for Ni$_2$FeGe has also been observed in previous calculations\,\cite{Kreiner:14} but contradicts to the expectations from Burch's rule as discussed by Kreiner et al. \cite{Kreiner:14} A more detailed study revealed that if we release the constraint of volume  conservation the L2$_1$ ordered phase becomes the ground state being 1.86 meV lower  in energy than the inverse one, see inset of  Fig.\,\ref{fig:coa} (b). However, since the energy difference is so small it is hard to determine which structure has to be expected experimentally a since this tiny energy barrier could be overcome at very low temperatures. So far the  Heusler structure was not observed for Ni$_2$FeGe. From recent high-temperature magnetocalorimetry experiments there is evidence that at high temperatures a cP4 (L1$_2$) structure can be stabilized.\cite{Yin:16} However, the Heusler structure might still be produced  in thin films. 
\begin{table*}
\renewcommand{\arraystretch}{1.1}
\begin{tabular}{C{1.2cm}C{1.5cm}C{1.6cm}C{1.5cm}C{1.5cm}C{1.5cm}C{1.5cm}C{1.5cm}C{1.5cm}C{1.5cm}}\hline\hline
e/a(Ni$_2$Y)&Y&&\multicolumn{7}{c}{\mbox{ }\hspace{.5cm}Z\mbox{ }\hspace{.5cm}}\\\hline
    &           &     &   B     &        Al&  Ga&      In&       Si&            Ge&      Sn\\\hline
27&Mn&here&1.38(r)\hspace{.25cm}\mbox{}&1.00(r)\hspace{.25cm}\mbox{}&1.25(r)\hspace{.25cm}\mbox{}&1.00(r)\hspace{.25cm}\mbox{}&1.00(r)\hspace{.25cm}\mbox{}&1.00(r)\hspace{.25cm}\mbox{}&1.00(r)\hspace{.25cm}\mbox{}\\
27    &          &other &1.38(r)\cite{Qawasmeh:12}&1.00(r)\cite{Enkovaara:03}&1.27(r)\cite{Qawasmeh:12}&&1.00(r)\cite{Fang:14}&1.18(r)\cite{Luo:13} &1.00(r)\cite{Krenke:05}\\
28&Fe&here&1.44(r)\hspace{.25cm}\mbox{}&1.35(r)\hspace{.25cm}\mbox{}&1.35(r)\hspace{.25cm}\mbox{}&1.00(r)\hspace{.25cm}\mbox{}& 1.36(r)\hspace{.25cm}\mbox{}&1.35(r)\hspace{.25cm}\mbox{}&1.00(r)\hspace{.25cm}\mbox{}\\
28 &                & other& &&1.35(r)\cite{Qawasmeh:12}&1.00(r)\cite{Luo:13} 1.30(r)\cite{Gillessen-Thesis}&&1.30(r)\cite{Gillessen:10} (i)\cite{Kreiner:14}&\\
29&Co& here&1.42(r)\hspace{.25cm}\mbox{}&1.38(r)\hspace{.25cm}\mbox{}&1.38(r)\hspace{.25cm}\mbox{}&1.35(r)\hspace{.25cm}\mbox{}& 1.30(i)\hspace{.25cm}\mbox{}& 1.30(i)\hspace{.25cm}\mbox{} & 1.30(i)\hspace{.25cm}\mbox{}     \\
29& &other&&1.40(r)\cite{Gillessen-Thesis}&1.38(r)\cite{Tavana:15}&$>$1.35\cite{Bai:12}&&1.30(r)\cite{Gillessen-Thesis} &1.30(r)\cite{Gillessen-Thesis}\\\hline
&e/a(Z)&&3&3&3&3&4&4&4\\\hline\hline
\end{tabular}
\caption{Calculated c/a ratios of the global minimum are given for  Ni-based Heusler alloys Ni$_2$YZ. Regular phases are denoted by r, the occurrence on the inverse structure is named i. For comparison value from literature has been added. e/a(Ni$_2$Y) and e/a(Z) give the number of valence  electrons for the metals  (Ni and Y) and the main group element on the Z site, respectively.}
\label{tab:coa}
\end{table*}
For Co containing alloys the symmetry depends on the choice of the Z element, i.e. Z elements from group IV prefer inverse order and the c/a ratio of the global minima is around 1.3 compared to 1.35 (1.38 for  Ni$_2$CoGa) for the L2$_1$ ordered alloys, cf. Fig.\,\ref{fig:coa}(c).  It should be noted that in case of Y = Co  the energy difference between the global minimum of the L2$_1$ and inverse  order can be analogous to the Fe case quite small (see Fig.~\ref{fig:emix} (a)), i.e. especially for Ni$_2$CoGa the difference is only  -12.55\,meV which corresponds to previous findings \cite{Dannenberg-Thesis} moreover the local minimum of the inverse structure becomes more stable than the one of the L2$_1$ ordered phase ($E^{inv}(c/a=0.92)-E^{L2_1}(c/a=0.865) = $37.12\,meV).

Moving from Mn via Fe to Co  the energy difference between the cubic and tetragonal phase increases from about 20\,meV for Ni$_2$MnGa to 200\,meV for Ni$_2$CoGa.
The Ni$_2$YB systems are exceptional because the difference in energy between the tetragonal and cubic phase is much larger, i.e. between 240 (Mn) and 400\, meV (Co) for the global minima and 20 to 150\,meV for the local minima with c/a $<$ 1. The difference of the total energies of  the high temperature austenite (cubic) phase and the low temperature martensite phase (here: tetragonal) can be viewed as a rough measure for the martensitic transition temperature. Barman showed that -- with certain restrictions -- the $\Delta E = E(c/a)_{\rm min}-E(c/a=1.0)$ basically proportional to the martensite temperature, i.e.  $k_{\rm B}T_{\rm M}$\cite{Barman:08}. Hence, the larger $\Delta E$ the higher $T_{\rm M}$ the  tetragonal  Ni$_2$CoZ alloys and Ni$_2$FeZ (Z = Ge, Ga, Si) phases should be more stable than  the well known Ni$_2$MnGa which transforms at 206 K to the cubic L2$_1$ structure\cite{Vasilev:99}.   Accordingly, the Fe and Co containing systems might be more interesting from the MAE point of view. However, one should keep in mind that relation between $\Delta E$ and $T_{\rm M}$ is a rough estimation and factors such as the number of valence electrons  $e/a$ and the structure (inverse or regular) have some impact.
Replacing one sublattice of Ni by Pd leads to a substantial shift of the c/a ratio to larger values (c/a = 1.42, see Tab.\ref{tab:nipd}) which is believed to improve the MAE. Besides the larger c/a ratio the energy difference  between the cubic and the tetragonal phase in NiPdFeGe is with 0.16\,eV/f.u.  3 times larger than the one of  the isoelectronic ternary Ni$_2$FeGe system which hints to a higher martensite temperature and a larger stability range of the tetragonal phase. 
\subsection{Phase stability}\label{sec:phase-stab}
In the previous section the stability of the L2$_1$ vs the inverse Heusler structure has been discussed. However,  so far not all investigated systems have been  synthesized and it is not known whether the observed ground state structure will be likely to stabilise these compounds experimentally or the system decomposes. In order to shed light on this the formation energy $E^{\rm form}$ has been investigated. For a given compound $E^{\rm form}$ can be obtained from 
\begin{equation}
E^{\rm form}\, = \,  E^{\rm Ni_2YZ}\,-\, (2E^{\rm Ni}\,+\,E^{\rm Y}\,+\,E^{\rm Z})\label{eq:Eform}
\end{equation}
\begin{figure}[bh]
\centering
\includegraphics[width=.99\columnwidth]{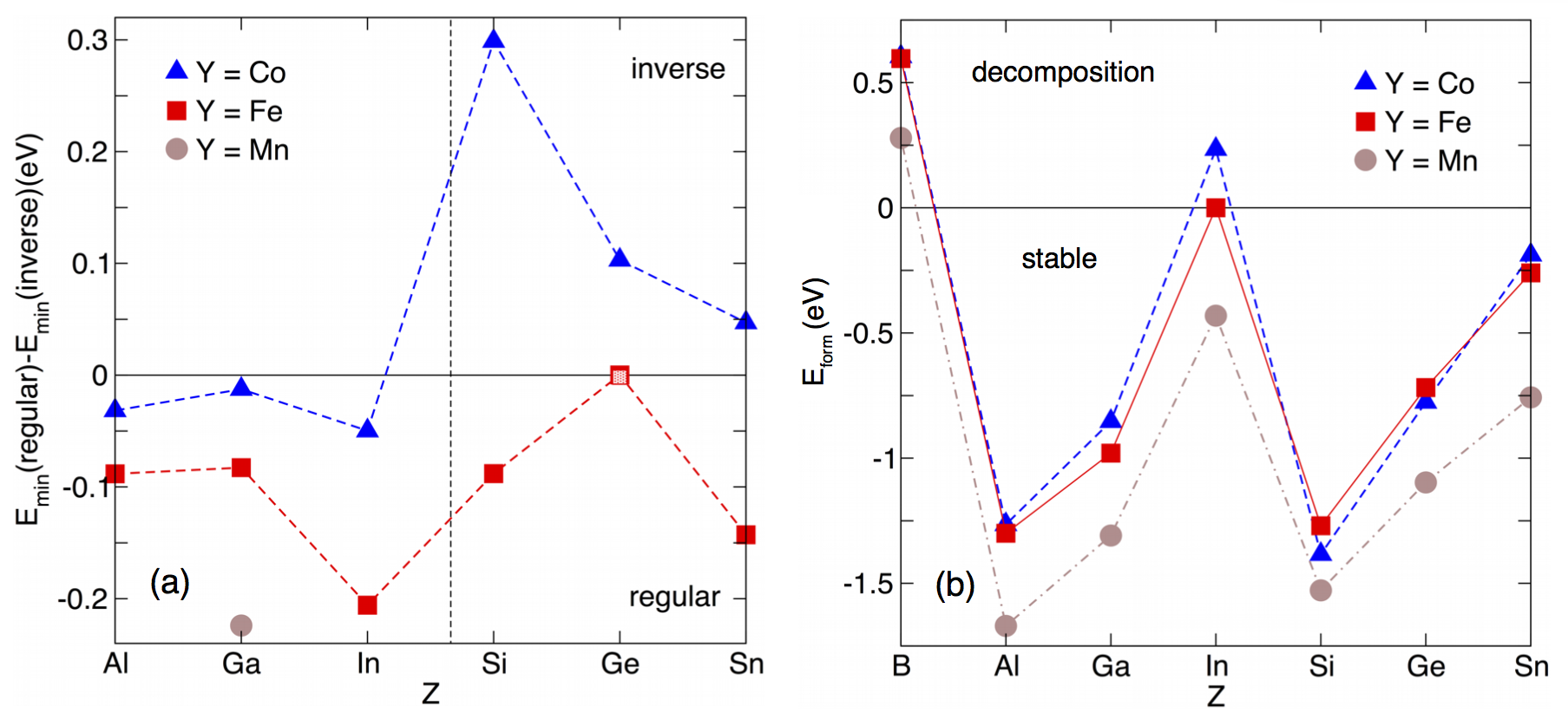}
\caption{(a) Calculated energy difference between the L2$_1$ and the inverse ordered phase for Ni$_2$FeZ (red squares) and Ni$_2$CoZ (blue triangles) compounds.  For the Mn compounds (brown circles) only the values for the systems with non cubic ground state are shown. (b) Energy of formation for Ni$_2$YZ Heusler compounds in with Y = Fe (red squares) and Co (blue triangles) as obtained from Eq.\,(\ref{eq:Eform}). Negative values denote systems stable against decomposition. Lines should only be viewed as guide to the eye.}
\label{fig:emix}
\end{figure}
where $E^{\rm Ni_2YZ}$ is the total energy at T = 0 \,K for the ground state configuration of the Ni$_2$YZ compound. The other terms $E^{\rm Ni}$, $E^{\rm Y}$, and $E^{\rm Z}$ correspond to the ground state energies calculated for the elemental systems. This can be viewed as a lower boundary and was chosen to handle all systems on the same footing. However, in case of stable binaries the formation energies might change and systems with  small negative formation energies might become unstable, too. No approximations have been made for the ground state structures of the elements, especially   $\alpha$-Mn  (I$\bar{4}$3m structure) and $\alpha$-B (experimentally observed structure taken from structure data base ICSD\,\cite{Bergerhoff:87,Belsky:02}) have been used as reference states. Huge differences in  the stability of the Heusler compound have been observed for the                                                                                                                                                                                                                                                                                                                                                                                                                             investigated systems depending on the Z element chosen, see Fig.\, \ref{fig:emix}(b). Independent from the element on the Y site all three series show the same trend,  
compounds with Al and Si are most stable whereas In and B containing compounds tend to decompose. In particular the B containing systems turned out to be not very likely to occur in nature  as a Heusler compound.  To our knowledge all studies  discussing N$_2$MnB for example in view of the high spin-polarization\cite{Qawasmeh:12} and  possible pressure behavior\,\cite{Pugaczowa-Michalska:08} are based on DFT investigations and  so far no  experimental verification has been found which agrees with the tendency to decompose observed in the present work. In case of In the trend is less unique.  Ni$_2$MnIn  is found to be stable with a formation energy of  -0.43\,eV whereas Ni$_2$CoIn decomposes ($E^{\rm form}$ = 0.23eV) and Fe is in-between with a formation energy being almost zero. In the last case  a reliable conclusion about the stability of this compound cannot be drawn. In case of Ni$_2$FeIn and Ni$_2$YSn the formation energies are small and the argument regarding elemental reference systems might apply. Indeed these are the systems of the least importance for this paper because they are cubic or provide very small MAE.
All other investigated systems turned out to be stable in the Heusler structure being consistent with previous theoretical findings.\cite{Kreiner:14,Gillessen:10} 
In the next section the magnetic properties are discussed focussing on the MAE of the stable systems with non-cubic ground states. 
\section{Magnetism}\label{sec:magn}
\subsection{Magnetocrystalline anisotropy of Ni$_2$YZ}
\begin{figure}[htb]
\centering
\includegraphics[width=.95\columnwidth]{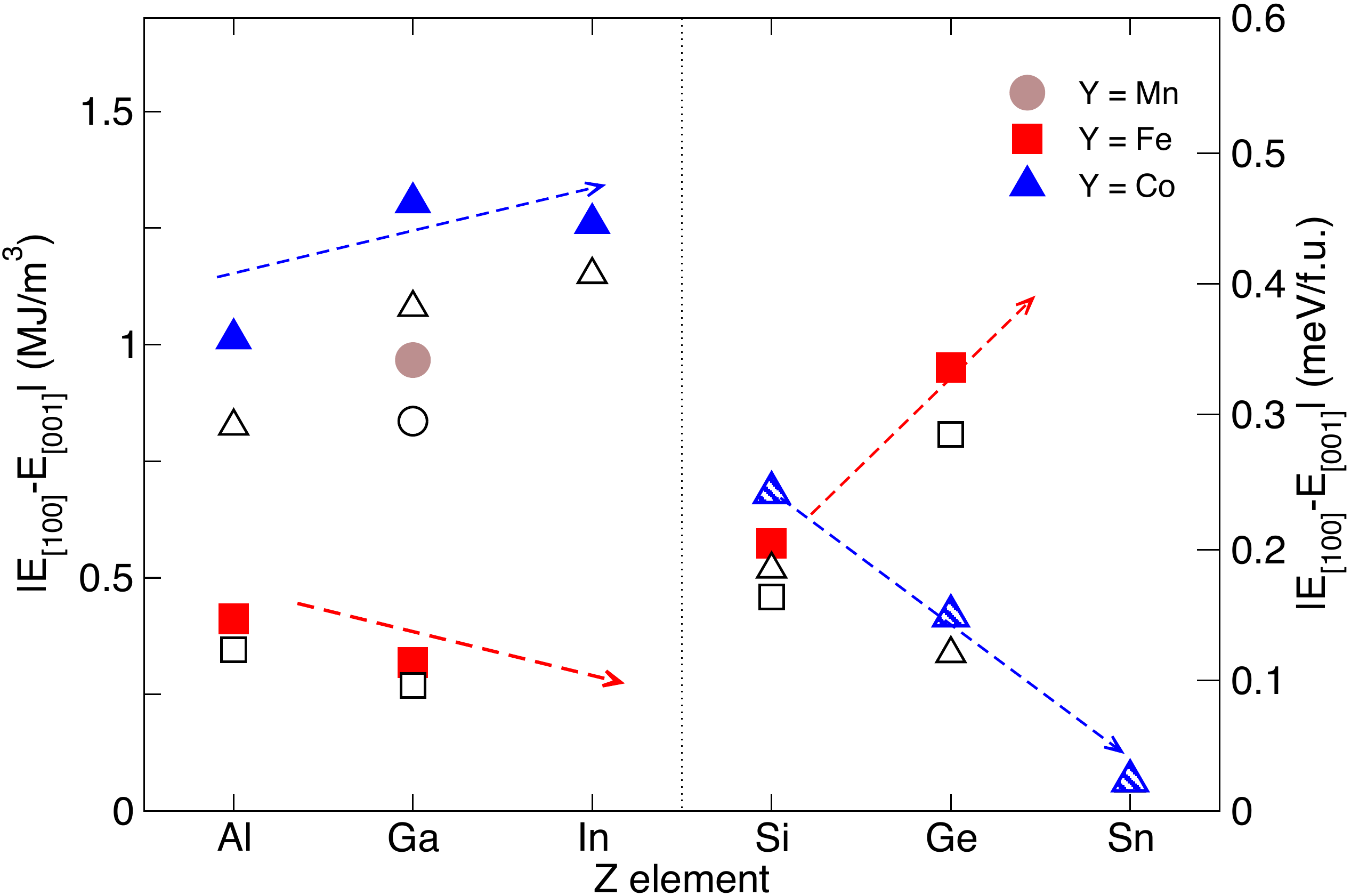}
\caption{Absolute values of the magnetic anisotropy energy per formula unit. The MAE in MJ/m$^3$ (scale on the left hand side)  is given by filled and hatched symbols where  filled (hatched) denotes regular (inverse) ordered Heusler systems. The open symbols show the MAE in atomic units e.g. meV per formula unit (scale on the right hand side).  For details see Sec.\,\ref{sec:methods}. Arrows mark the overall trends.}
\label{fig:mae-ea}
\end{figure}

We have performed highly accurate MAE calculations within the full potential LMTO method using the optimized structures from previous VASP calculations. Although the investigated compounds have   similar c/a ratios, number of valence electrons, and comparable electronic structure, the spread in the MAE values turned out to be quite big. For the ground state configurations with c/a$>$1 it reaches from  about 0.42 \,meV/f.u.  to almost zero, see Fig.~\ref{fig:mae-ea}. The highest MAE values are  achieved for systems with Y=Co and  a Z element from group III. In case of Ni$_2$CoGa and Ni$_2$CoIn MAE values of  1.30\,MJ/m$^3$ (0.38\,meV/f.u.) and  1.26\,MJ/m$^3$(0.42\,meV/f.u). The MAE for the two above mentioned Y = Co compounds  is about 30\% larger than  the value obtained for the well known  Ni$_2$MnGa system  which shows an MAE value of about 0.34\,meV/f.u., see Fig.\ref{fig:mae-ea}.  Gruner {\em et al.} observed an even larger value for c/a $>$ 1.25 (MAE = 0.6 meV/f.u.) in Ni$_2$MnGa~\cite{Gruner:08} using the fully relativistic minimum basis set approach FLPO~ in the local density approximation.\cite{Gruner:08} 
For the Ni$_2$MnGa system numerous experimental studies exist which have studied the anisotropy. The values spread from  $K_u = 1.17\cdot10^6$~erg/cm$^3$  (0.117 MJ/m$^3$) to 4-5$\cdot10^6$~erg/cm$^3$\ (0.4-0.5 MJ/m$^3$),\cite{Ullakko:96,Albertini:01,Sakon:13} depending on the exact composition, and whether single or multi-variant samples have been used. In addition, temperature effects play a role and can reduce  the size of the magneto crystalline anisotropy\cite{Albertini:01}. Our calculated anisotropy energy for Ni$_2$MnGa is of the same order of magnitude as the experimental values for single crystals~\cite{Tickle:99}  by a factor of 3-4 larger which is a quite good agreement regarding the fact that we consider an ideal system at zero Kelvin. Furthermore, the comparison to calculated values in other works show a spreading in the theoretical data too for example due to the potential used \,\cite{Gruner:08}. The calculated values depend partially on the volume, the  computational method and the approximations made but  they give the right order of magnitude. Though the absolute values are sensitive to the computational method the sign is in our cases robust and it is possible to obtain trends within a series of compounds or alloys calculated on the same footing which can be a guideline for the design of new materials with even larger MAEs. 

While Co has been proven to be advantageous if Z is taken from main group III (Z = Al, Ga, In) Ni$_2$CoZ compounds with  Z being Si, Ge, or Sn are less promising. The calculated  MAE  values are in agreement with Ref.\,\onlinecite{Faleev:17} significantly smaller, e.g.  0.42\,MJ/m$^3$ in case of Ni$_2$CoGe.  In contrast to Ni$_2$CoGa(In) the easy axis is out-of plane. However, even though these systems are uniaxial ferromagnets the magnetic properties are less good due to the inverse order. One reason is the change in the local coordination of the magnetic Co atom which has 4 Ni and 4 Z atoms instead on 8 Ni atoms, see Sec.\,\ref{sec:magn}. Therefore, the magnetization of the inverse ordered systems is smaller compared to the L2$_1$ type systems.  In addition, this effects the lattice structure and leads to smaller $c/a$ values for the ground state configuration.

Taking everything into consideration i.e. MAE, magnetic moments, and phase stability  Ni$_2$FeZ (Z = Si, Ge) seem to be advantageous. Both systems provide a  tetragonal ground state and an easy axis MAE. The largest MAE occurs for Ni$_2$FeGe  0.95\,MJ/m$^3$, see Fig.~\ref{fig:mae-ea}. This is comparable to the findings for the  Mn-based tetragonal Heusler systems.~\cite{Winterlik:12} Since for Ni$_2$FeGe the L2$_1$ and inverse ordered structure are very close in energy (see discussion in Sec.\,\ref{sec:phase}) the MAE of the inverse ordered system has also been investigated. It turned out to be  by a factor of 2 smaller then the one of the regular Heusler compound (c/a= 1.3, MAE=0.54\,MJ/m$^3$). This observation is in line with the findings for the inverse ordered Ni$_2$CoZ compounds. Summarizing, the largest values for the MAE have been predicted for Ni$_2$CoZ (Z = Ga, In) and Ni$2$FeGe which have the same number of valence electrons per f.u. (e/a = 32). 
 
From the c/a variation discussed in Sec.\ref{sec:phase} it is obvious the tetragonal distorted systems have not only a global minimum at $c/a > 1$ but also a local minimum for compressed systems with $c/a < 1$. In some cases the global minimum  is not reached and the local minimum  $c/a <  1$  appears. The MAE values  for the local minima at $c/a < 1$ tend to be smaller  what is at least partially related to the smaller deviation from the cubic structure. For example in case of Ni$_2$CoGa, the system with the highest MAE for $c/a > 1$, the MAE for  the local minimum  ($c/a=$ 0.87) is with -0.10\,MJ/m$^3$ by a factor of 10 smaller, see Fig.\,\ref{fig:mae-coa}. On the other hand  moving from elongation to compression is accompanied by a change from easy plane to easy axis anisotropy. This has been reported before for Ni-based Heusler compounds such as Ni$_2$MnGa\,\cite{Gruner:08} and Ni$_2$Mn$_{1.25}$In$_{0.75}$ \,\cite{Krumme:15}. For Ni$_2$MnGa a MAE of 0.186\,meV/f.u. (c/a = 0.94) is achieved which is in good agreement with the full potential augmented plane wave calculations by Enkovaara {\em et al.} (MAE = 0.18 meV/f.u.).\cite{Enkovaara:02}
Due to the smaller tetragonal distortion the size of the MAE falls usually behind the one of the previously discussed case for  $c/a > 1$. An exception is Ni$_2$FeGe (MAE(c/a=0.85)=0.20\,meV) system where the change is comparably smaller. This underlines that the c/a ratio is not the only determining factor.  No or very tiny MAE values are obtained for the inverse ordered Ni$_2$CoZ systems at $c/a<1$ since the magnetic moment  vanishes with decreasing c/a (Z = Si, Ge) and the tetragonal distortion of the local minimum is small (Z = Sn, c/a= 0.92, MAE = -0.004~meV), see Supplement. 

\subsection{Possible routes to tailor the magnetocrystalline anisotropy}
It is fair to say that some  Ni-based Heusler compounds reveal promising MAEs, but the most interesting systems (in view of large MAE values) have positive formation energies, are easy-plane systems, or could not be synthesized as bulk systems even if predicted  by theory to be stable.  Here we focus on possible ways how to tailor the size and the sign of the MAE and simultaneously stabilize the tetragonal phase. Special focus will be on the effect of lattice deformation, forming quaternary compounds to turn easy-plane magnets into easy-axis systems. 
Knowing that heavier materials provide a larger spin-orbit coupling are are therefore likely to possess also a larger anisotropy then 3d materials we consider also the influence of 4d element replacements of Ni. In the present paper we have studied in particular Pd which is isoelectronic to Ni.
\subsubsection{Changing the lattice geometry}
\begin{figure}[htb]    
\centering
\includegraphics[width=.9\columnwidth]{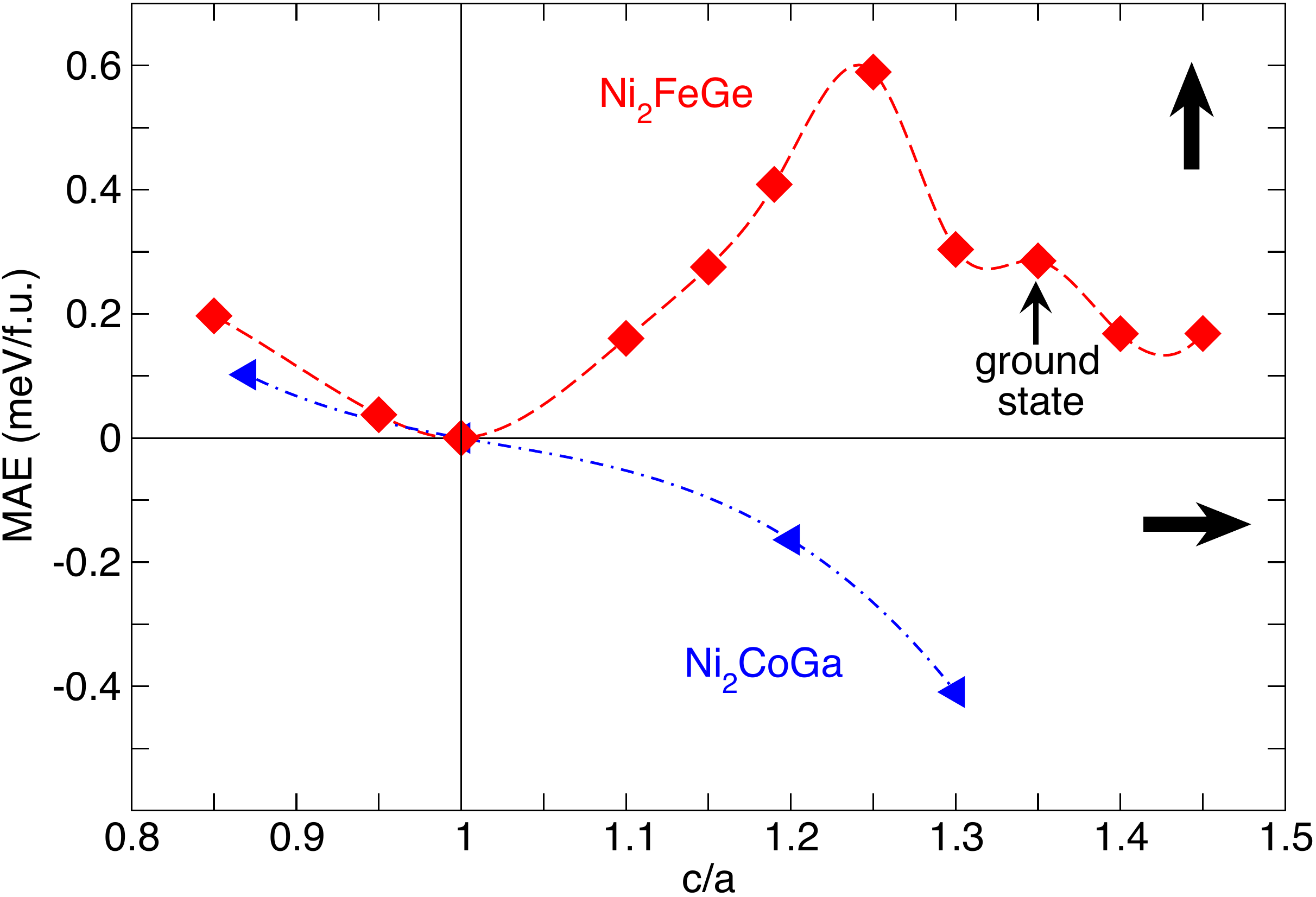}
\caption{Dependence of the magneto crystalline anisotropy on the lattice ratio c/a for Ni$_2$FeGe  (red diamonds) and Ni$_2$CoGa (blue triangles) as obtained from RSPt. In case of  Ni$_2$FeGe no sign change i.e. rotation of the MAE from easy axis to planar is observed depending on the c/a ratio whereas Ni$_2$CoGa undergoes the typical change from planar (negative values) to uniaxial when the c/a changes from $c/a>1$ to $c/a<1$.
\label{fig:mae-coa}}
\end{figure}
Changing the lattice distortion from elongation to compression of the $c$ axis is in  many Heusler type compounds  accompanied by a rotation of the easy axis from in-plane to out-of-plane orientation.\cite{Krumme:15, Gruner:08} The variation is almost linear with $c/a$ in some cases flattening around the minima, one example is shown in Fig.~\ref{fig:mae-coa} for Ni$_2$CoGa. another example for an off-stoichiometric Ni-based Heusler system can be found  in Ref.\,\onlinecite{Krumme:15}. Such a change in the orientation of the easy axis is not observed for the systems  which possess an out-of-plane axis for $c/a>1$,  see supplement and Fig.\ref{fig:mae-coa}. A strong non-linear $c/a$ dependence  occurs in case of  Ni$_2$FeGe. Reducing the tetragonal distortion by  about 10\%  ($c/a = 1.25$)  leads to a 100\% increase of  the MAE compared to the ground state ($c/a = 1.35$), cf Fig.\,\ref{fig:mae-coa}. 
With 1.96 MJ/m$^3$ the MAE is even larger than the values obtained for Ni$_2$CoZ and what is even more important the system is uniaxial. The large MAE and  the possibility to tune  by small lattice distortions  might make this system interesting  in view of applications. However, to our knowledge it has not been successfully synthesized as a bulk system in L2$_1$ structure. However, one might think to stabilize a structure as a thin film where the c/a ratio could be optimized by dopants or deposition or the choice of the substrate. 
\begin{figure}[htb]
\centering
\includegraphics[width=.95\columnwidth]{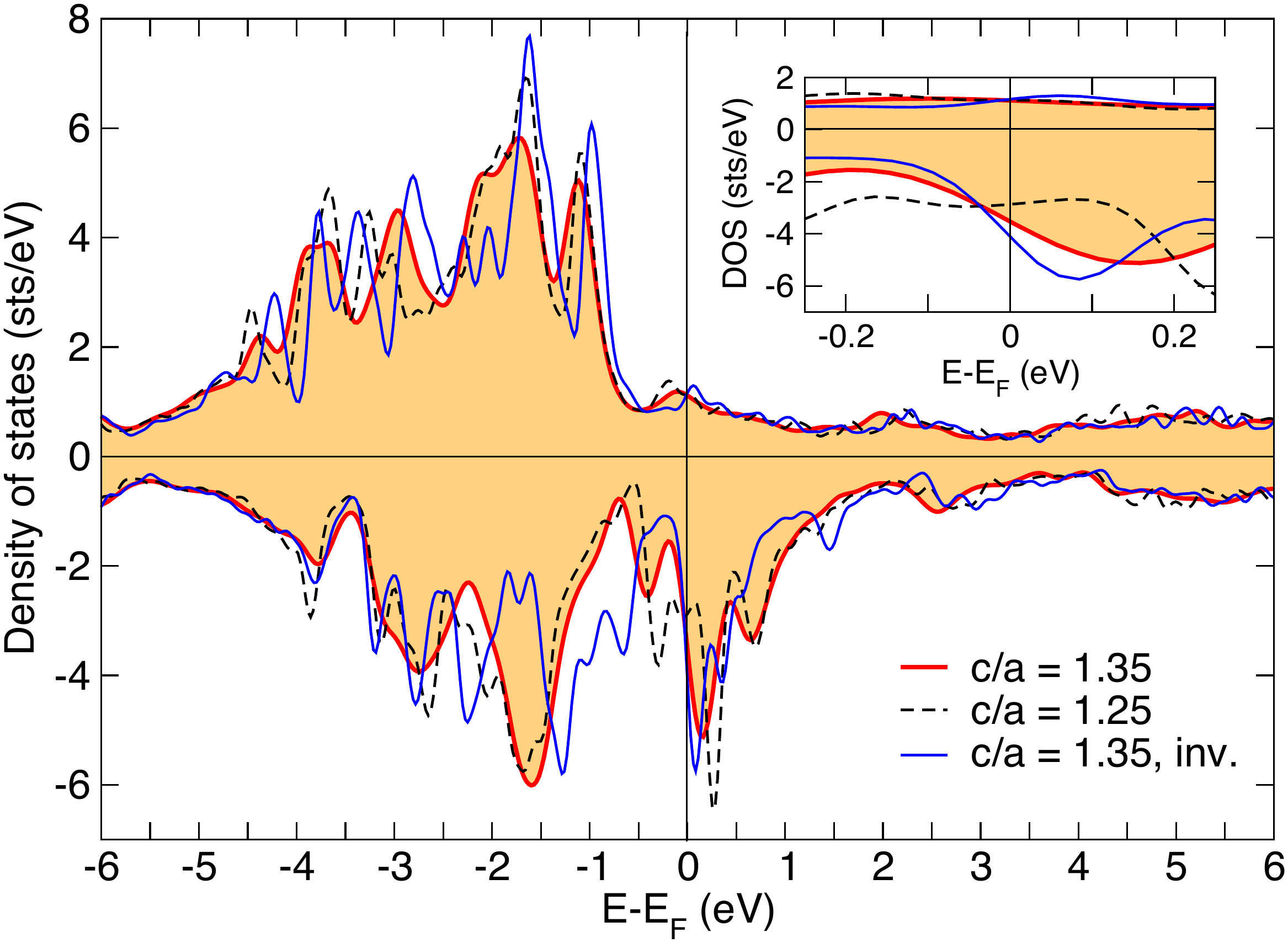}
\caption{(a) Calculated total density of states of Ni$_2$FeGe in the ground state with c/a = 1.35 (red line, shaded area), for the configuration with the largest MAE (c/a =1.25) (dashed line), and the global minimum of the Hg$_2$CuTi ordered structure with c/a = 1.35 (thin blue line). The inset shows the region around the Fermi level (vertical line). Here the minority-spin DOS at $E_{\rm F}$ decreases from the inverse ordered structure over the global minimum to the artificially distorted structure with c/a=1.25 whereas the MAE increases at the same time.
\label{fig:dos-nfg-tot}}
\end{figure}
The remaining question is what drives the changes in the magnetic behavior and how is it related to changes in the electronic structure? Comparing the  total density of states (DOS)  of the tetragonal ground state (c/a = 1.35) to the  DOS of the  squeezed structure (c/a = 1.25) and the  ground state (c/a = 1.3) of the inverse ordered system characteristic changes of the DOS  at the Fermi level can be observed. Hence, the DOS at the Fermi level is dominated by the  3d states of the transition metals (see also  Fig.\,\ref{fig:dos-d}) the differences in Fig.\,\ref{fig:dos-nfg-tot} reflect the changes in the 3d states. Reducing the c/a ratio leads here to an increase of 2.4\%  (decrease of 0.9\%) of the Fe-Fe (Fe-Ni) distance and a decrease of the DOS in the minority channel at the Fermi level. Hence, the DOS decreases from the inverse structure (c/a$_i$ = 1.35) over the ground state of the regular structure (c/a$_r$ =1.35) to the squeezed (c/a$_r$ = 1.25) structure  the MAE of these three systems behaves just opposite, i.e. it increases from  (c/a$_i$ = 1.35) which has a negligible MAE of about 0.09 MJ/m$^3$ over the ground state (about 1MJ/m$^3$) to the distorted configuration with a twice as large MAE. It should be noted that despite the huge changes in size the MAE remains always uniaxial for this system.
\subsubsection{Quaternary compounds}

\begin{figure}[bht]
\centering
\includegraphics[width=.98\columnwidth]{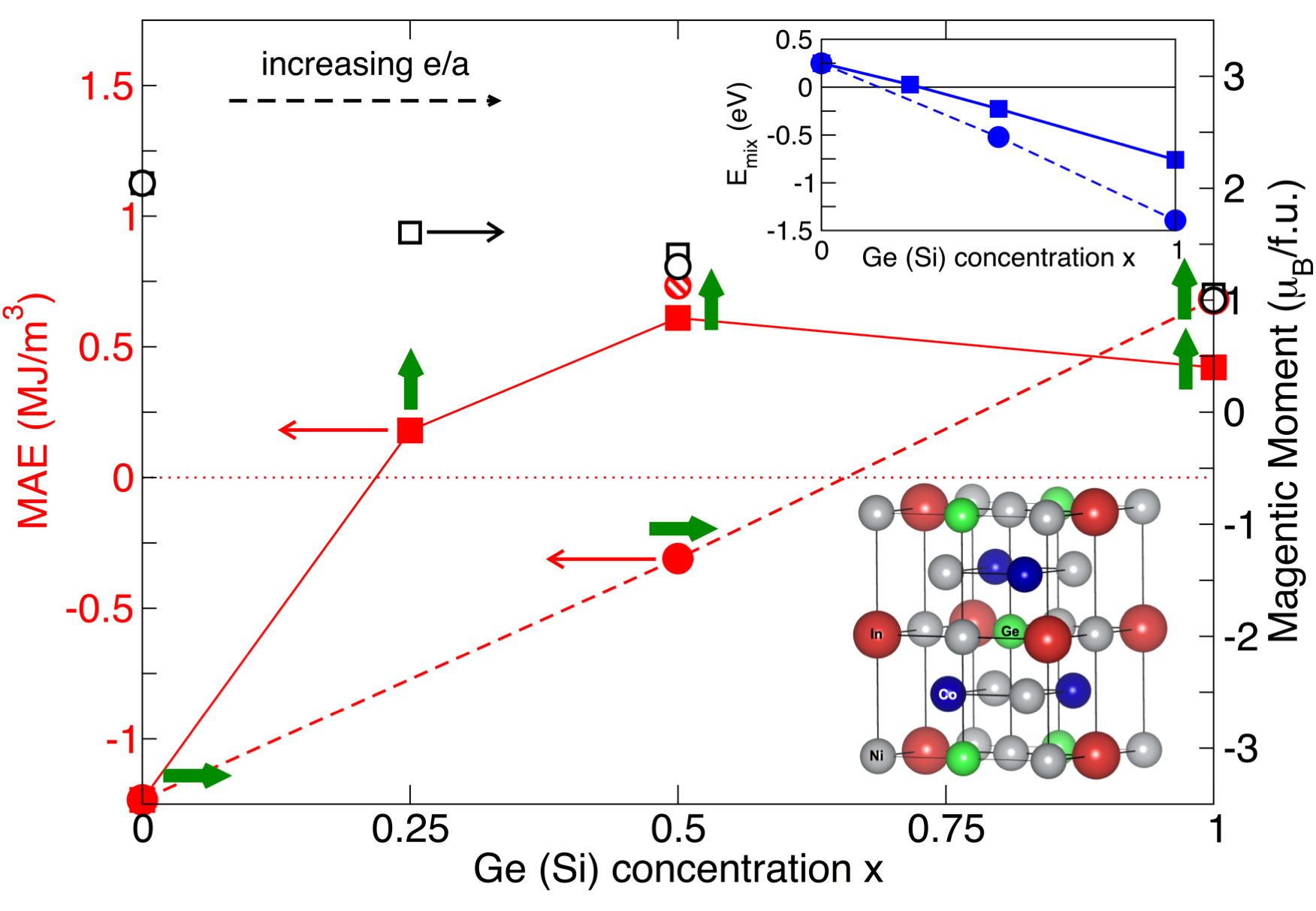}
\caption{Calculated dependence of the magneto crystalline anisotropy on the impurity concentration i.e. the valence electron number e/a for Ni$_2$CoIn$_{1-x}$Z$_x$ with Z = Ge (squares) and Si (circles).  The open black symbols denote the corresponding magnetic moments per formula unit (f.u.), see scale on the right hand side. The inset at the top right  shows the formation energy and  stabilization of the system with increasing Ge and Si content, respectively. The inset on the bottom right hand side shows the super cell (16 atoms) used for the quaternary systems, here  50\% of In have been substituted by Ge. The hatched circle corresponds to the MAE of the local minima of Ni$_2$CoIn$_{0.5}$Si$_{0.5}$ (c/a = 1.2). The thick (green) arrows indicate the orientation of the MAE.
\label{fig:mae-inge}}
\end{figure}

While in the previous section structural changes within one system were discussed we will focus here on alloying on one sublattice to improve the magnetic properties. Two examples have been chosen. In the first case the Z sublattice is used to improve the MAE of Ni$_2$CoZ. On the one hand from Fig.\,\ref{fig:mae-ea} it is conclusive  that  Ni$_2$CoGa and Ni$_2$CoIn have the largest MAE values, unfortunately is not an uniaxial anisotropy and as discussed in previously the  In compound is  not expected to be stable at low temperatures (cf Sec.\ref{sec:phase}). On the other hand the inverse ordered Ni$_2$CoGe has a much smaller MAE but it is uniaxial and the compound is according to our survey stable in the Hg$_2$CuTi  structure. Therefore, it seems a natural choice to replace In partially by Ge which means increasing the e/a. Here, In has been chosen over Ga because of the larger atomic number the  spin-orbit ($LS$) coupling is expected to be larger and this in turn should counteract  the reduction which is expected due to Ge. We replaced  25 and 50\% of  In  by Ge using a 16 atomic super cell, see Sec.\,\ref{sec:methods}.  The quaternary  Ni$_2$Co(In$_{0.5}$Ge$_{0.5}$) compound turns out to be stable (cf inset of Fig.\,\ref{fig:mae-inge}) and as for Ni$_2$CoIn the inverse ordered structure is lower in energy.   Compared to Ni$_2$CoIn or  Ni$_2$CoGa  the MAE is with 0.61\,MJ/m$^3$ by a factor of  2 smaller (50\% Ge),  but what is more important is  that the partial replacement of In by Ge has led to an easy axis anisotropy, see Fig.\,\ref{fig:mae-inge}. Smaller Ge concentrations would give smaller MAE values, but the magnetic anisotropy remains uniaxial as in the original compound Ni$_2$CoIn. Furthermore, for Ge concentrations equal to 25\% or less the quaternary compound decomposes, see inset in Fig.\,\ref{fig:mae-inge}. Summarizing, by  mixing Ni$_2$CoGe and Ni$_2$CoIn one gets basically the best of both systems, i.e. a stable phase with uniaxial MAE. However, the magnetic moment of the quaternary compound is reduced compared to Ni$_2$CoIn which is a tribute to the inverse ordered structure in which the moments are smaller compared to regular Heusler compounds, see also Sec.\,\ref{sec:magn}.  
Though, replacing In partially by Ge stabilizes the compound and improves the magnetic properties i.e. turns the system in an uniaxial magnet similar behavior might be expected by using Si instead of Ge since the preconditions are very similar but the ternary Ni$_2$CoSi compound shows an even larger uniaxial MAE. However, substituting 50\% of In by Si leads to an inverse ordered tetragonal structure with c/a$<1$, namely c/a = 0.905 and a local minimum at c/a = 1.2 (see supplement). In contrast to the previous case with Si the MAE remains planar if 50\% of Ge are replaced by Si, see Fig.~\ref{fig:mae-inge}.  The MAE changes quasi linearly from Ni$_2$CoIn to Ni$_2$CoSi leading to a planar MAE of 0.31 MJ/m$^3$ for Ni$_2$CoIn$_{0.5}$Si$_{0.5}$. This difference between Si and Ge  seems to be related to the change of the neighbor distances in the quenched phase (c/a$<$1.0), hence for the local minimum (c/a = 1.2) the situation is the same as in the In case. The MAE is uniaxial being slightly larger than for the ternary parent system Ni$_2$CoSi, see hatched circle in Fig.\ref{fig:mae-inge}. 

In the second case we followed the same line of argument but tailoring the occupation of the Y sublattice instead. In this case we keep the Z element fixed. In our case we have chosen Z = Al. Both ternary compounds are regular Heusler systems with a tetragonal ground state and are stable according to Fig.\,\ref{fig:emix}. Aiming to increase the MAE of  Ni$_2$FeAl  Fe has been partially replaced by Co, since Ni$_2$CoAl has a larger MAE which is  unfortunately planar, see Fig.\,\ref{fig:mae-ea}. However, instead of improving the magnetic properties as in the previous example the MAE almost vanishes. remains For  Ni$_2$Fe$_{0.75}$Co$_{0.25}$Al calculated in a 16 atom super cell the MAE remains uniaxial but decreases to a value $< 0.1$MJ/m$^3$.  

\subsubsection{Isoelectronic replacement of Ni}
It has been  pointed out in Bruno's model\,\cite{Bruno:89} the MAE is directly related to the spin-orbit coupling strength $\xi$ of a system and  knowing that $\xi$ is related to the atomic number by $\xi \sim Z^4$, i.e. heavy elements are preferable aiming for a large MAE. Therefore we replaced Ni partially by Pd  assuming that  an isoelectronic exchange will  improve  the magnetic properties, especially the MAE and leave the other properties such as the phase stability unchanged. As test system we have chosen Ni$_2$FeGe which has already a suitable MAE of about 1 MJ/m$^3$. Half of the Ni atoms  in this compound have been replaced by Pd  and calculations have been performed using a 16 atomic supercell. For comparison we have also studied the Ni-free system Pd$_2$FeGe. As expected the isoelectronic replacement has no significant influence on the phase stability. The formation energies are with -0.676 eV (NiPd)$_2$FeGe and -0.881\,eV (Pd$_2$FeGe) in the same range as for Ni$_2$FeGe, see stars in Fig.\,\ref{fig:emix}. Inducing Pd in the Heusler compound to replace Ni should have only minor influence on the electronic structure (isoelectronic) and preserve to uniaxial MAE. This is indeed observed the MAE remain uniaxial and the volume increases due to the larger Pd atom, see Tab.\,\ref{tab:nipd}. Replacing 50\% of the Ni atoms (16 atom super cell) leads to an increase of the volume per f.u. by  12\% but unfortunately the MAE does not change much. It basically remains constant at 0.97 MJ/m$^3$, cf Tab.\,\ref{tab:nipd} but would increase the prize by a factor of 1400. So the replacement of Ni by Pd would not only be inefficient but also incredibly expensive. 
\begin{table}[htp]
\caption{Structure data and magnetic properties for a series of isoelectronic Heusler compounds Ni$_{1-x}$Pd$_x$FeGe. Note the magnetic moments shown here are the total magnetic moments including the orbital moment as obtained from full potential DFT calculations using RSPt\,\cite{RSPt}. Positive sign for the MAE indicates uniaxial anisotropy.}
\begin{center}
\begin{tabular}{C{1.9cm}|C{1.1cm} C{1.6cm}C{1.6cm}C{1.6cm}}\hline\hline
System &c/a& Volume & m$_{\rm tot}$ & MAE \\
 && (\AA$^3$/f.u.)& ($\mu_{\rm B}$/f.u.) & (MJ/m$^3$)\\\hline
Ni$_2$FeGe& 1.35&48.00 &3.50&0.95\\
(NiPd)$_2$FeGe& 1.42& 53.82&3.40&0.97\\
Pd$_2$FeGe& 1.38&58.86& 3.24&0.61\\\hline\hline
\end{tabular}
\end{center}
\label{tab:nipd}
\end{table}%
\subsection{Magnetic moments}\label{sec:moment}
\begin{figure}[tbh]
\centering
\includegraphics[width=.9\columnwidth]{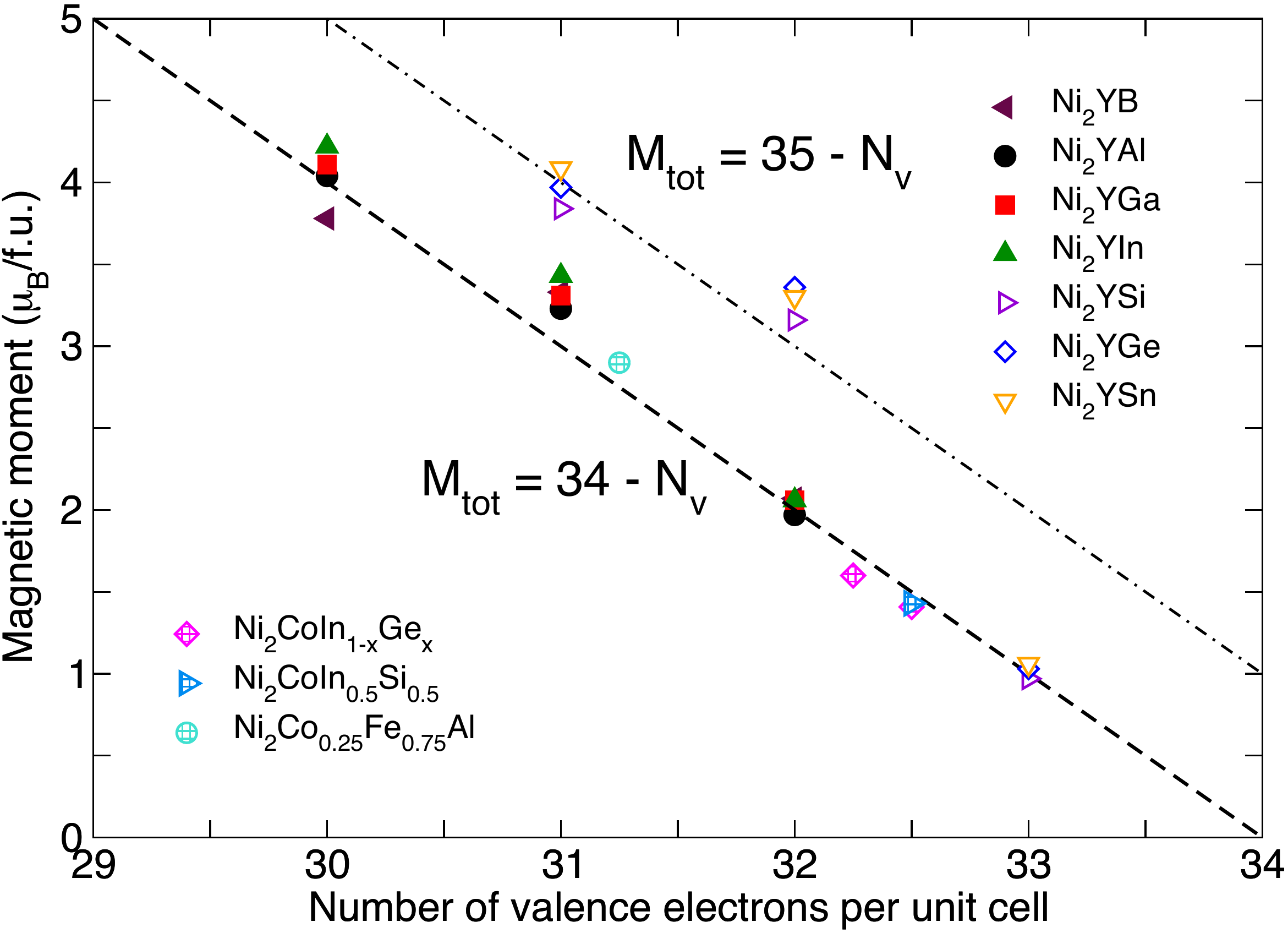}
\caption{Total magnetic moments of the ground state. The value of the cubic (c/a = 1.0) structure is given in brackets. All values are in $\mu_{\rm B}$/f.u.  The dashed line marks the 34-$N_v$ rule for shape memory alloys as suggested by Dannenberg {\em et al.}\cite{Dannenberg-Thesis} whereas the magnetic moment for the inverse ordered systems with the  Z element being from group IV follows shows a 35-$N_v$ behavior (dashed-dotted line). Inverse ordered systems including the quaternary ones (hatched symbols) follow the dashed line since their moment is reduced due to the reversed nn positions.}
\label{fig:magn}
\end{figure}
\begin{figure}[htb]
\centering
\includegraphics[width=.9\columnwidth]{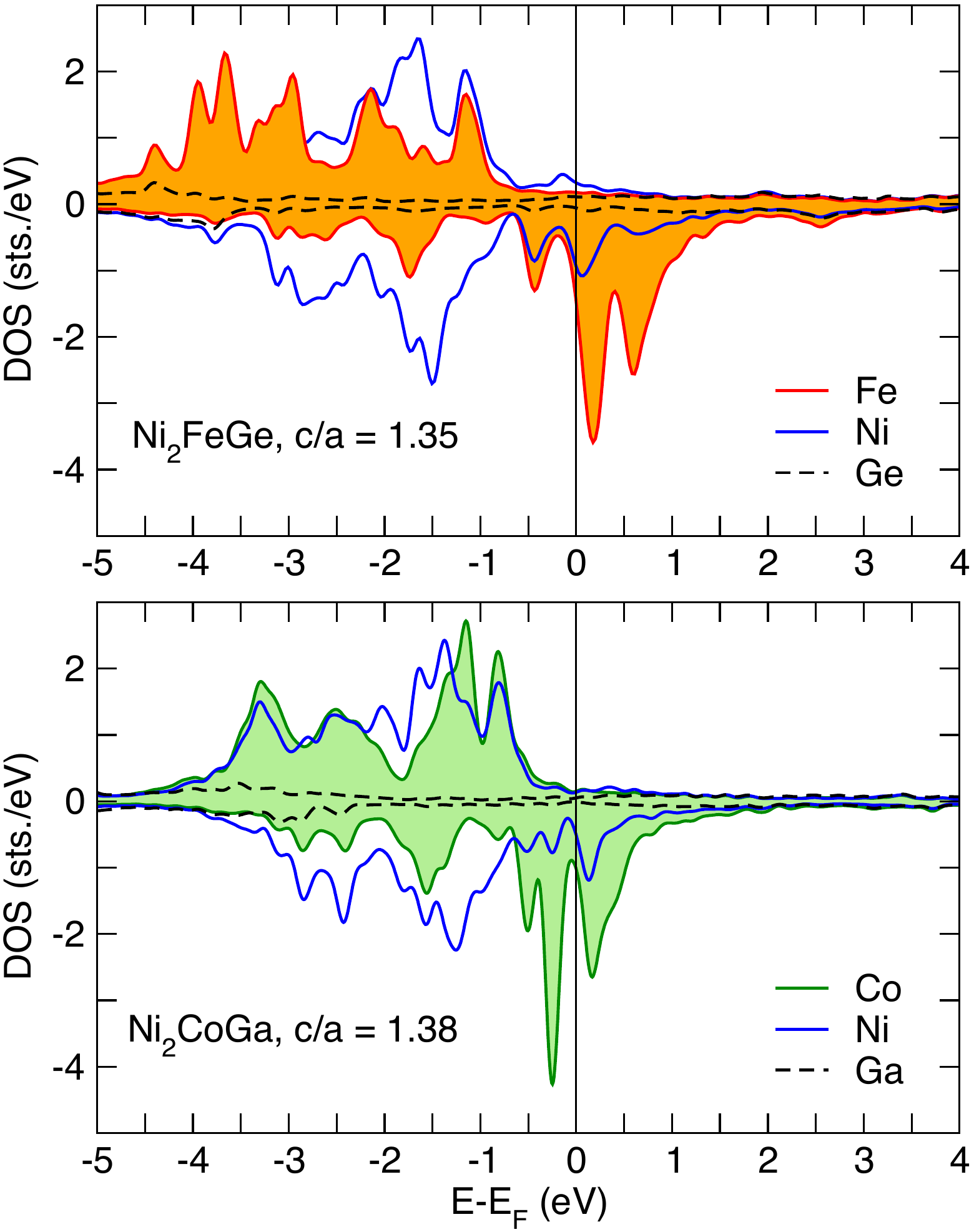}
\caption{Orbital resolved density of states for Ni$_2$FeGe (a) and Ni$_2$CoGa (b). Shaded areas denote the Fe and Co 3d states, respectively. The blue solid line marks the 3d DOS of Ni and the Ge(Ga) p-states are shown as dashed lines.
\label{fig:dos-d}}
\end{figure}
The magnetic moments  of Heusler alloys show usually  Slater-Pauling-type behavior, i.e.  their total magnetic moments depend linearly on the number of valence electrons. First demonstrated for  L2$_1$ ordered Co based half metallic ferromagnets  similar behavior has been observed  for related systems. In case of the halfmetallic Co-based compounds the magnetic moments are given by  $M = N_v$-24 with $N_v$ being the number of valence electrons per formula unit\cite{Galanakis:04} which gives integer magnetic moments for stoichiometric ordered systems.  It has turned out that  this rule can be generalized for  many  classes of Heusler alloys. Half-metallic Heusler alloys X$_2$YZ with X being an early 3d transition metal obey depending on the Y and Z constituent slightly different rules namely  $M= N_v  - 18$ and $M = N_v - 28$ \cite{Skaftouros:13}.  For Ni$_2$Mn$_{1-x}$Ga$_x$ alloys Dannenberg proposed a $M = 34 - N_v$ behavior of the total magnetic moment.\cite{Dannenberg-Thesis} As shown in Fig.~\ref{fig:magn} also all Ni-based Heusler alloys with a Z element from main group III obey this rule. Adding electrons to the system by occupying the Z sublattice with an element from main group IV increases the spin moments such that they follow the rule $M = 35 - N_v$. However, one should keep in mind that not the number of valence electrons  $N_v$ decides which rule the system obeys but the choice of the Z element. Taking for example  Ni$_2$CoGa and Ni$_2$FeGe, both systems have $N_v = 32$ but obey different rules, i.e. the Fe compound has a higher magnetic moment, see Fig.\,\ref{fig:magn}. This related to the fact that the DOS at the Fermi level is mostly determined by the 3d states of the transition metals, cf Fig.\,\ref{fig:dos-d}. In case of Ni$_2$CoGa the minority spin channel more occupied and therefore the spin moment smaller compared to the Fe compound.  A peculiarity occurs for the inverse ordered Ni$_2$CoZ with Z an element from group IV. We have  argued in the previous sections that inverse ordered Heusler compounds for the same type of compound have smaller magnetic moments due to different local order. This observation can be quantified as shown in Fig.\,\ref{fig:magn}. These compounds obey the $M = 34 - N_v$ rule instead. The same holds for the quaternary compounds with inverse lattice structure, see hatched symbols in Fig.\,\ref{fig:magn}.
\section{Conclusion}\label{sec:conclusion}
Ni-based Heusler compounds Ni$_2$YZ have been used as an example to study different routes to improve the magnetic properties with special focus on the MAE. The magnetic moments of the systems follow modified Slater-Pauling laws. Showing clearly that the magnetic moment of the inverse ordered systems is systematically lower than for regular Heusler compounds. 

Out of the 21 studied systems 14 possess a non-cubic ground state being a prerequisite for a finite MAE. From these candidate systems the ones with Co on the Y sublattice turned out to be most interesting. In Ni$_2$CoZ compounds  the tetragonal phase turned out to be most stable, i.e. the transformation to the cubic austenite phase will occur at higher temperature as for example for Ni$_2$MnGa. However, the Ni$_2$CoZ (Z= In, Ga) systems with the largest MAE turned out  to have a planar MAE and/or  are even unstable. We could show that this could be cured by combining them  with  inverse ordered Ni$_2$CoGe which has a uniaxial MAE. Using Si instead of Ge turned out not to be successful, because the tetragonal phase  in Ni$_2$CoSi  has c/a $< 1 $. For the local minimum at c/a = 1.2 the same effect as for Ge is observed. Hence, the phase  can be stabilized by adding valence electrons  (replacing In partially by Si or Ge) but to improve the magnetic properties also the lattice structure of the ternary phases has to match. Aiming to  increase the uniaxial MAE of Ni$_2$FeAl we used the same strategy, but replacing partially Fe by Co since Ni$_2$CoAl as a larger (planar) MAE. However, it turned out that 25\% Co of the Fe sublattice  reduce  the MAE drastically. This is not completely unexpected since changing the coordination and magnetism can lead to a reduction of the MAE.  

Another way to improve the MAE could be the use of heavier elements, which possess larger spin orbit coupling. To test this for our set of systems we selected the ternary system with the largest uniaxial MAE, Ni$_2$FeGe. Partial isoelectronic replacement of Ni by Pd showed only minor effect on the MAE, because due to Pd the volume increases and the magnetic moments slightly decrease both facts counteract to an increase of the MAE. 

The MAE also changes with the lattice ratio such that one can think to tailor the MAE by stress or strain. Basically this  c/a dependence has been discussed for Ni$_2$MnGa and other systems in literature before. Most systems show a quasi linear behavior with a sign change at c/a =1. However, in case of Ni$_2$FeGe the MAE remains uniaxial for all lattice ratios between 0.85 and 1.45. For this particular system small deviations from the equilibrium c/a boost the MAE from  about 1 to 2 MJ/m$^3$. Similar behavior is expected for other  Ni$_2$FeZ compounds since the orientation of the MAE for the ground state and the local minimum at $c/a<1.0$ are the same.

Concluding for  the systems under consideration we discussed different routes to manipulate the MAE e.g. creating quaternary compound by doping or by lattice deformation. Both routes turned out to be successful under certain conditions and MAE values of 1 MJ/m$^3$ and higher could be achieved.  Limitations for practical use are the height of martensite temperature and the Curie temperature. Especially the Curie temperature has recently be predicted to be quite low in some of the systems. However,  though we have limited ourselves to Ni-based Heusler compounds we believe that the results can carry over to other Heusler compounds and with this also the finite temperature properties might be improved.
\section*{Acknowledgements}
HCH would like to thank Miroslav Werwinski and Yaroslav Kvashnin for helpful discussions and support.  We acknowledge financial support  by the EU through the Horizon 2020 framework programme NOVAMAG (No. 686056) and STandUP for energy. Computational resources were provided by the Swedish National Infrastructure for Computing (SNIC).

\end{document}